\documentclass[12pt,a4paper]{article}
\usepackage{amsmath,amssymb,mathrsfs,framed,esint,slashed,tikz,appendix}
\usetikzlibrary{arrows.meta,positioning,calc}
\usepackage[colorlinks]{hyperref}
\usepackage{color}
\usepackage[all]{xy}
\emergencystretch=\maxdimen
\newcommand{\rem}[1]{}

\newcommand{\beq}{\begin{equation}}
\newcommand{\eeq}{\end{equation}}
\newcommand{\bal}{\begin{align}}
\newcommand{\eal}{\end{align}}

\numberwithin{equation}{section}

\begin{document}

\title{\vspace{-.7cm}Euler-Poincar\'e Formulation of Barotropic Fluids Coupled with ADM Gravity}
\author{Allan Louie$^{1}$ \smallskip 
\\
\footnotesize
\it $^1$Blackett Laboratory, Imperial College, Prince Consort Road, London, SW7 2AZ, UK
\\
\footnotesize
}
\date{$\,$}
\maketitle

\vspace{-1.4cm}
\begin{abstract} \footnotesize
This paper develops a geometric mechanics framework for the reduction of general relativistic hydrodynamic variational principles, from the variation of worldlines approach in 4D spacetime to 3-dimensional Eulerian descriptions. We consider a self-gravitating, barotropic fluid and obtain the Euler-Poincar\'e equations of the system by Lagrangian reduction. Using the decomposition of general relativity into \(3+1\) dimensions, with a direction of time defined, the gauge invariance of the action over spacetime diffeomorphisms permits a 3-dimensional description of the fluid diffeomorphism by gauge fixing. The configuration space thus mirrors the Newtonian case, and by employing the Euler-Poincar\'e theorem, we derive the Eulerian equations of motion, in the same form as the PDEs from Newtonian fluid dynamics. The equations of motion are then derived, where the fluid variables are measured in a separate frame of reference from the variables involving gravity. Furthermore, the general relativistic barotropic fluid exhibits a Kelvin-Noether circulation conservation, which we derive in both the inertial and moving frames of reference. Finally, potential applications to Numerical Relativity are discussed.

\end{abstract}

\vspace{-.6cm}
{\scriptsize}

\section{Introduction}

Variational principles appear throughout nature. First proposed by Lagrange in 1760, they remain succinct formalisms that contain dominant information on the dynamics of a system represented only by its kinematical and geometric information. Much of their appeal arises from their suitability for analysis, regarding Hamiltonian formulations, structural hierarchies in the resulting equations of motion, symmetries, and conservation laws.

A leading model concerning general relativistic fluid modelling is the Pull-back Approach, first proposed by Taub \cite{taub_1954_general}.  In addition to Einstein's equations, it postulates that the collection of particle worldlines, modeled using a map from the space of fluid labels to spacetime, must minimise the spacetime volume spanned. This model is preferred by mathematicians and theorists as it shares the benefits of a variational principle and may be extended to model multifluid systems. While insightful and dominant in theory, the resulting Euler-Lagrange equations of motion are difficult to simulate because the fluid map itself belongs to the space of all possible embeddings of the fluid reference labels into spacetime. 

This simulation aspect invites some variant of Euler-Poincar\'e reduction \cite{holm_1998_the}, which reduces abstract Lagrangian mechanics over Lie groups down to partial differential equations for the Eulerian velocity field while retaining the variational structure of the parent theory. An approach that parallels Euler-Poincar\'e reduction for this covariant field theory has been done in \cite{gaybalmaz_2024_general}, where the Eulerian velocity equation in 4 dimensions was derived.

However, simulating field equations covariantly is still a difficult task. We note that Einstein's equations for general relativity, which are 10 coupled, second-order partial differential equations over spacetime, are difficult to simulate due to challenges regarding computational cost and numerical stability. A common method to tackle this is with the \(3+1\) formalism, decomposing the spacetime continuum into one temporal and three spatial components and making this a Cauchy or initial value problem. The first model, the ADM (named after Arnowitt, Deser an Misner) formalism, was introduced in 1962 \cite{arnowitt_2008_republication}, providing a basis for symplectic integration and constraint handling. Though superseded by modern approaches such as CCZ4 \cite{alic_2012_conformal} or BSSN \cite{baumgarte_1998_numerical, shibata_1995_evolution}, the philosophy of a \(3+1\) split remains. This will be the approach taken in this paper, where both the metric tensor and hydrodynamics will be described using the \(3+1\) formalism.

This paper aims to illustrate the geometric insight retained within the \(3+1\) formalism of relativistic hydrodynamics, despite performing a splitting of spacetime and losing covariance. It also aims to reveal the inherited variational structure regarding those partial differential equations by deriving them from a covariant action from theoretical gravitational physics, to enhance the potential for future analysis by connecting relativistic hydrodynamics with Newtonian fluid dynamics. The paper does not aim to propose a direct alternative for integrating the Eintein-Euler equations, as optimised formalisms such as \cite{mart_1991_numerical} build upon hyperbolicity of the equations, which we do not analyse in this paper.

Specifically, the paper performs the \(3+1\) decomposition for the Pull-back Action for barotropic fluids on both the metric and fluid variables to obtain 3-dimensional Euler's equations coupled with ADM gravity. While the \(3+1\) decomposition of the metric variables reproduces the familiar ADM formalism, it is shown in this paper that the decomposition simplifies the kinematics of the fluid map so that Euler-Poincar\'e reduction may be directly applied, placing the equations on equal footing with nonrelativistic fluid dynamics. Furthermore, the variational structure of the model provides a coordinate-free description of the system, which we utilise here by deriving the same equations in a time-dependent frame of reference. Finally, we derive the Kelvin circulation theorem for the system, which indicates the preservation of the Euler-Poincar\'e structure \cite{holm_1998_the} within the system.

\subsection{Contents of the Paper}

This paper obtains the Euler-Poincar\'e equations for ideal fluids coupled with ADM gravity by performing the \(3+1\) split and Lagrangian reduction on Taub's covariant action. The plan is as follows: Section \ref{Section 2} discusses the Pull-back formalism, along with a recap of its symmetries. Section \ref{Section 3} shows that the \(3+1\) split of spacetime permits a reduction scheme for the fluid action, allowing an expression in terms of Eulerian variables only. Section \ref{Section 4} derives the stress-energy tensor of the fluid, and reveals its Euler-Poincar\'e nature before deriving the fluid momentum equation by reduction. Section \ref{Section 5} expresses the Lagrangian in terms of moving frame variables and derives the corresponding Euler equation. Section \ref{Section 6} derives the corresponding Kelvin circulation theorem in both the inertial and moving frames. Finally, we conclude with a summary and discussion of results, and an outlook for future research.

\section{Pull-back Formalism} \label{Section 2}

In this section, we discuss the results we need to provide a geometric setup for what follows. Following a similar construction to \cite{andersson_2021_relativistic}, the Pull-back formalism, a covariant variational principle for fluid worldlines in curved spacetime, is discussed below.

Let spacetime \(\mathbf{M}\) be a simply connected, Lorentzian 4-dimensional manifold without boundaries, with metric \(g\in T_2^0(\mathbf{M}) \), and let \((M,\tilde{n})\) be a 3-dimensional manifold endowed with a volume form representing fluid labels and the reference number density, respectively. The trivial bundle \(\pi:W=M \times \mathbb{R} \to M\) represents the world-history label of the fluid. A fluid map is a smooth embedding, denoted

\begin{equation}
    \Psi \in Emb(W,\mathbf{M}),
\end{equation}
so \(Emb(W,\mathbf{M})\) is the configuration space for fluids.

The vector field \(J\in \mathfrak{X} (\mathbf{M})\), representing the number density current of the fluid, is defined by:

\begin{equation} \label{J2.2}
    \Psi^* \iota_{J} \epsilon_g=\pi^* \tilde{n}.
\end{equation}
\(\epsilon_g \in \Lambda^4(\mathbf{M)}\)is the volume form associated with \(g\). In coordinates, \(\epsilon_g=\sqrt{-g}\ d^4x\) using the \((-,+,+,+)\) signature.

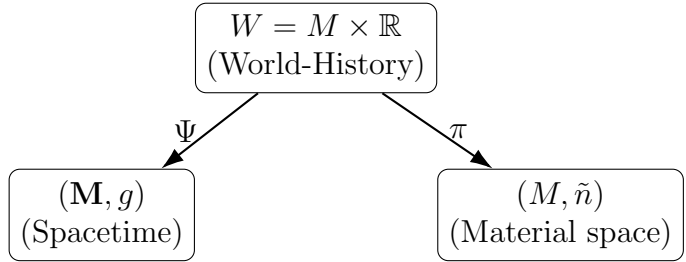
\begin{figure}[t]
\centering
\begin{tikzpicture}[
  box/.style={draw, rounded corners, inner sep=4pt, align=center},
  arr/.style={-{Latex[length=3mm,width=2mm]}, thick}
]

\node[box] (W) {$W=M\times \mathbb{R}$\\(World-History)};
\node[box, below left=10mm and 0mm of W] (Sp) {$(\mathbf{M},g)$\\(Spacetime)};
\node[box, below right=10mm and 0mm of W] (Mat) {$(M,\tilde{n})$\\(Material space)};

\draw[arr] (W) -- node[left] {$\Psi$} (Sp);
\draw[arr] (W) -- node[right] {$\pi$} (Mat);

\end{tikzpicture}
\caption{\footnotesize Schematic representation of the Pull-back formalism of relativistic hydrodynamics. The world-history \(W\) contains material space \((M)\) and time \((\mathbb{R})\) labels of the barotropic fluid parcels, and the time label is projected out using the trivial bundle \(\pi:W\to M\). The fluid labels are mapped onto physical spacetime \(\mathbf{M}\) by the embedding \(\Psi\in Emb(W,\mathbf{M})\).}
\label{fig1}
\end{figure}

The above geometric setup may be illustrated using a mindmap, which is shown in Figure \ref{fig1}.

Finally, by defining

\begin{equation}
    n=\sqrt{-g(J,J)} \in C^\infty (\mathbf{M})
\end{equation}
as the baryonic number density of the fluid, the Pull-back Action for barotropic fluids may be defined\footnote{The geometric construction of this paper follows Andersson and Comer \cite{andersson_2021_relativistic}, but the energy density depends on \(n\) rather than \(n^2\) in this paper. This is a choice of convention and does not affect the physics, and is used in Taub \cite{taub_1954_general}, Carter \cite{carter_1973_elastic} and Holm \cite{holm_1985_hamiltonian}.} as follows:
\begin{equation} \label{taubpullback}
    S[g,\Psi]= \int_\mathbf{M} \epsilon_g (^{(4)}R-\rho(n)),
\end{equation}
where \(^{(4)}R\) is the Ricci Scalar on \(\mathbf{M}\), and \(\rho\) is a real-valued function, representing the energy density of the fluid. In this paper, the analysis is done on the simplest case of barotropic fluids to exemplify the analysis and techniques used to convert this theoretical framework to obtain compressible fluid equations. However, this may be extended to models for thermal, charged, or multifluid systems (eg. GRMHD) through the inclusion of additional field variables or multiple fluid maps.

The first term of \eqref{taubpullback} is recognized as the Einstein-Hilbert action, while the second term represents the world-volume of the fluid, the continuum analog to the geodesic action for discrete particles.

The action has two gauge symmetries, which we outline below.

\subsection{Spacetime Diffeomorphisms}

Consider the spacetime gauge transformation 

\begin{equation}
    \Psi\to \Phi \circ \Psi, \ g\to(\Phi^{-1})^* g,
\end{equation}
under \(\Phi \in Diff(\mathbf{M})\).

The transformation of the vector field \(J\in \mathfrak{X}(\mathbf{M})\) is derived below.

\begin{align*}
(\Phi \circ \Psi)^* \iota_{J'} \epsilon_{(\Phi^{-1})^*g} &=\pi^* \tilde{n}\\
(\Phi \circ \Psi)^* \iota_{J'} \epsilon_{(\Phi^{-1})^*g} &=\Psi^* \iota_J \epsilon_g\\
\Psi^*\Phi^* \iota_{J'} \epsilon_{(\Phi^{-1})^*g} &=\Psi^* \iota_J \epsilon_g\\
\Psi^* \iota_{(\Phi^{-1})_*J'} \epsilon_g &=\Psi^* \iota_J \epsilon_g\\
(\Phi^{-1})_*J'&=J\\
J'&=\Phi_*J.
\end{align*}

Thus, \(J\to\Phi_*J \) transforms as the push-forward under \(\Phi\), and \(n\) transforms as a scalar field. \(\epsilon_g\), and thus the product \(\rho(n)\epsilon_g\) transforms as a volume form. Therefore, the action is invariant under spacetime diffeomorphisms.

\subsection{Material Diffeomorphisms}

Consider the material gauge transformation

\begin{equation}
    \Psi\to\Psi\circ w,\ g\to g,
\end{equation}
with \(w\in Aut(W)\), satisfying \(\pi \circ w=\bar{w} \circ \pi \) for some \(\bar{w}\in Diff(M)\).

The transformation for \(J\) by \(w\) is as follows.

\begin{align*}
(\Psi\circ w)^* \iota_{J'} \epsilon_g&=\pi^* \tilde{n}\\
w^* \Psi^* \iota_{J'} \epsilon_g&=\pi^* \tilde{n}\\
\Psi^* \iota_{J'} \epsilon_g&=(w^{-1})^*\pi^* \tilde{n}\\
\Psi^* \iota_{J'} \epsilon_g&=\pi^*(\bar{w}^{-1})^* \tilde{n}.\\
\end{align*}

The transformation of \(J\) by \(w\) may not leave the action invariant. Thus, the gauge symmetry concerning material diffeomorphisms is \(Aut(W)_{\tilde{n}}\), the isotropy subgroup of \(\tilde{n}\), for which \(J\to J\). This broken symmetry implies the semidirect-product action of 
\(Aut(W)\) on the co-set \(\Gamma := Aut(W)/Aut(W)_{\tilde{n}}\) \cite{holm_1998_the}. 

\section{ADM Decomposition and Reduction to 3-Dimensional Fluid Dynamics} \label{Section 3}

In the ADM formalism of general relativity, spacetime is foliated by the association \(\mathbf{M}\cong W\). From here onward, we denote spacetime as \(W\), and \(\Psi\in \textit{Diff}(W)\) as a smooth invertible map on \(W\).

Given that spacetime possesses the trivial bundle structure of \(W\), the metric may be decomposed into ADM variables as follows:

\begin{equation} \label{ADMsplit}
    g=(\beta^a \beta_a -\alpha ^2)dt^2+\beta_a (dx^adt+dtdx^a)+\gamma_{ab}dx^adx^b,
\end{equation}
and the associated volume form is 

\begin{equation} \label{epsilong}
    \sqrt{-g}=\alpha\sqrt{\gamma}.
\end{equation}

Here, \(\alpha\) is named the lapse function, \(\beta^a\) are the components of the shift vector, and \(\gamma_{ab}\) are the components of the spatial metric. Similar to the role of the covariant metric in 4D general relativity, \(\gamma_{ab}\) is used to lower indices in the ADM formalism, while its inverse, \(\gamma^{ab}\), is used to raise indices. For example, \(\beta_a=\gamma_{ab} \beta^b\). Modulo boundary terms, the Einstein-Hilbert action on this trivialised spacetime is

\begin{equation} \label{SADM}
    S_{ADM}=\int dt \int_M d^3x \alpha \sqrt{\gamma} (K_{ab}K^{ab}-K^2 +R),
\end{equation}
where \(K_{ab}\) are the components of the extrinsic curvature, \(K=K^a_a\) and \(R\) is the Ricci scalar on \(M\). Detailed descriptions of the ADM formalism are found in \cite{arnowitt_2008_republication, misner_1995_gravitation}.

It is now shown that the ADM trivialisation of spacetime also permits a \(3+1\)-dimensional split to the fluid variables, allowing the components of \(J\) to be written explicitly in terms of a trivialised fluid map, so that the Euler-Poincar\'e theorem may be applied directly. An authoritative summary of Lagrangian reduction on general relativistic fluids is found in \cite{gaybalmaz_2024_general}.

We note that it is always possible to choose a gauge \(\Phi\in \textit{Diff}(W)\) in which  \(\Psi'=\Phi\circ \Psi\) preserves the time variable under the transformation. This means \(\Psi'(x,t)=(x',t)\) for some \(x'\in M\), for any \((x,t)\in W\). The spacetime diffeomorphism gauge invariance highlighted in Section \ref{Section 2} then states that the action in terms of \(\Psi'\) takes the same form as \(\Psi\). Hence, by fixing a gauge, all possible fluid maps may be written in the form:

\begin{equation}
    \Psi(x,t)=(h_t(x),t),
\end{equation}
for all \(x\in M\), and \(h_t\in \textit{Diff}(M)\ \forall t\in \mathbb{R}\). Together with the ADM split of the metric, the covariant field-theoretic action on \(W\) may be interpreted as Lagrangian mechanics of fluid parcels and 3-geometries on \(M\). This may be written as:

\begin{equation} \label{S}
    S[\Psi,g] = S[h_t,\dot{h}_t, \gamma_{ab}, \dot{\gamma_{ab}},\alpha,\beta^a ].
\end{equation}

The left-hand side of \eqref{S} is the field-theoretic action in terms of the covariant fluid map and metric, while the right-hand side decomposes the fluid map into the gauge-fixed fluid map \((h_t, \dot{h}_t)\), and the metric into the ADM variables \((\gamma_{ab}, \dot{\gamma_{ab}},\alpha,\beta^a)\). It is noted that the action on the right-hand side does not depend on the rate of change of the lapse and shift \((\dot{\alpha},\dot{\beta^a})\). The action on the right-hand side of \eqref{S} is thus specified by taking an integral curve over the following manifold, in the above order:

\begin{equation}
    L: T(\textit{Diff}(M) \times T^0_2(M)) \times C^\infty(M) \times\mathfrak{X}(M) \to \mathbb{R},
\end{equation}
with\footnote{The assumption of a boundaryless spacetime requires time to be integrated in the interval \((-\infty,\infty)\) in the Einstein-Hilbert action. This means there is no Noether term for any continuous symmetry in this system.}

\begin{equation}
    S=\int Ldt.
\end{equation}

The reduction of the parcel Lagrangian to Eulerian variables is performed by expressing \(J\) in terms of \((h_t, \dot{h}_t)\), which is done in Appendix \ref{Appendix B}. The result is:

\begin{equation} \label{J^0}
    J^0=\frac{\tilde{n}_0}{\det(\nabla h_t)\alpha \sqrt{\gamma}},
\end{equation}
and 

\begin{equation} \label{J^a}
    J^a  \partial_a=J^0 \dot{h}_th_t^{-1}, \ a=\{1,2,3\},
\end{equation}
where \(\tilde{n}_0\) is defined as the coefficient of the fluid label volume form, so that \(\tilde{n}=\tilde{n}_0\ d^3x\). We note that equation \eqref{J^a} relates vector fields on spacetime, denoted \(\mathfrak{X}(W)\), with \(\{\partial_t,\partial_1, \partial_2, \partial_3\}\) forming a basis.

As we will see in Section \ref{Section 4}, the Eulerian picture of the Pull-back Action requires \(J^0\) and \(J^a\) to be interpreted according to their transformation rules under the action of \((h_t, \dot{h}_t)\). Here, we note that \(J^0\) transforms as an Eulerian number density, the coefficient to a volume form \(\Lambda^3(M)\). Then, we decompose \(J=J^\mu \partial_\mu\) as follows:

\begin{equation}
    J^0 u:=J=J^0(\dot{h}_th_t^{-1},1).
\end{equation}

The spatial components \(u^a=(\dot{h}_th^{-1}_t)^a\) may also be interpreted as the Eulerian fluid velocity associated with \((h_t, \dot{h}_t)\), and is a spatial vector field \(\mathfrak{X}(M)\).

It is noted that the above decomposition of \(J=:J^0u\) differs from the conventional \(J=:nU\) as seen in \cite{andersson_2021_relativistic}, which normalises the Eulerian velocity by \(-g(U,U)=1\). As will be shown in the next section, this alternative decomposition allows the separated variables to be viewed as central quantities in the Eulerian picture.

This section concludes by expressing the fluid part of the Pull-back Lagrangian explicitly in terms of \((h_t, \dot{h}_t)\).

The number density is:

\begin{align*}
n&=\sqrt{-g(J,J)}\\
&=J^0\sqrt{-g(u,u)}\\
&=J^0\sqrt{-(\beta^a \beta_a -\alpha^2)(u^0)^2-\beta_a (u^au^0+u^0u^a)-\gamma_{ab}u^au^b}\\
&=J^0\sqrt{-(\beta^a \beta_a -\alpha^2)-2\beta_au^a-\gamma_{ab}u^au^b}\\
&=J^0\sqrt{\alpha^2-\gamma_{ab}(\beta^a+u^a)(\beta^b+u^b)}.\\
\end{align*}

Hence, the fluid action is:

\begin{equation} \label{Sfluid}
    S_{fluid}=-\int_W d^4x\ \alpha \sqrt{\gamma}\rho \left(J^0\sqrt{\alpha^2-\gamma_{ab}(\beta^a+u^a)(\beta^b+u^b)}\right),
\end{equation}
using the expression in \eqref{epsilong} for the volume element. It is noted that the expression inside the square root is similar to the kinetic energy of ideal fluids, but with the shift vector \(\beta^a\) added to the Eulerian velocity, and with the lapse function providing an offset.
\section{Equations of Motion of the Decomposed ADM-Fluid Lagrangian} \label{Section 4}

Recalling that the Lagrangian kinematically depends on the manifold \(T(\textit{Diff}(M) \times T^0_2(M)) \times C^\infty(M) \times\mathfrak{X}(M)\), the extremisation of the action yields an equation of motion for each variable. In this section, a reduction procedure will be applied to part of the configuration space. In particular, the \((\gamma_{ab},\dot{\gamma_{ab}},\alpha,\beta^a)\) and \((h_t, \dot{h}_t)\) equations are derived separately, with the first half providing information on how matter curves spacetime, and the second encodes the dynamics of fluids under the influence of the spacetime metric.

\subsection{Euler-Lagrange Equations of the ADM Variables}

\subsubsection{Metric evolution equation}

Summing the expressions in \eqref{SADM} and \eqref{Sfluid}, the Pull-back Action written in terms of the ADM variables and the trivialised fluid map is:

\begin{equation} \label{S3+1}
    S= \int dt \int_M d^3x\ \alpha \sqrt{\gamma} \Bigg(K_{ab}K^{ab}-K^2 +R-\rho \left(J^0\sqrt{\alpha^2-\gamma_{ab}(\beta^a+u^a)(\beta^b+u^b)}\right)\Bigg).
\end{equation}

The \(\rho\) term is associated with the matter term in general relativity. The Euler-Lagrange equation for \(\delta \gamma_{ab}\) is the same as the covariant formulation of general relativity \cite{misner_1995_gravitation}:

\begin{equation}
    G^{ab}=\frac{1}{2}T^{ab}
\end{equation}
where in this case, the Einstein tensor corresponds to the variation of the ADM action with respect to the spatial metric,

\begin{equation}
    G^{ab}=-\frac{1}{\alpha \sqrt{\gamma}}\frac{\delta S_{ADM}}{\delta \gamma_{ab}},
\end{equation}
and the stress-energy tensor corresponding to the fluid action is:

\begin{equation}
    T^{ab}=\frac{2}{\alpha \sqrt{\gamma}}\frac{\delta S_{fluid}}{\delta \gamma_{ab}}.
\end{equation}

By making use of this result, finding the Euler-Lagrange equations for metric evolution becomes a problem of finding the stress-energy tensor, which we compute below.

Firstly, we note that the fluid Lagrangian does not depend on the time derivatives of the ADM variables, so that:

\begin{equation}
    \frac{\delta L_{fluid}}{\delta \dot{\gamma}_{ab}}=0
\end{equation}

As a result,

\begin{align*}
\frac{\delta S_{fluid}}{\delta \gamma_{ab}}&=\frac{\delta L_{fluid}}{\delta \gamma_{ab}}\\
&=-\alpha\frac{\delta \sqrt{\gamma}}{\delta \gamma_{ab}}\rho-\alpha \sqrt{\gamma}\frac{\delta \rho}{\delta \gamma_{ab}}\\
&=-\alpha\Big(\frac{1}{2}\sqrt{\gamma}\gamma^{ab}\Big)\rho-\alpha \sqrt{\gamma}\frac{d\rho}{dn}\frac{\delta n}{\delta \gamma_{ab}}\\
&=-\frac{1}{2}\alpha \sqrt{\gamma}\rho \gamma^{ab}-\alpha \sqrt{\gamma}\frac{d\rho}{dn}\Big(-\frac{1}{2}n\Big)\left((\beta^a+u^a)(\beta^b+u^b)\frac{(J^0)^2}{\alpha^2}+\gamma^{ab}\right)\\
&=\frac{1}{2}\alpha \sqrt{\gamma}\Big(n\frac{d\rho}{dn}-\rho\Big) \gamma^{ab}+\frac{1}{2}\frac{(J^0)^2}{n}\alpha \sqrt{\gamma}\frac{d\rho}{dn}(\beta^a+u^a)(\beta^b+u^b)\\
\end{align*}

By using the thermodynamic equation of state for pressure \(p=n\frac{d\rho}{dn}-\rho\) \cite{misner_1995_gravitation}, we have:

\begin{equation}
    T^{ab}=p\gamma^{ab}+\frac{(J^0)^2}{n}\frac{d\rho}{dn}(\beta^a+u^a)(\beta^b+u^b)
\end{equation}

This formula may be compared with the standard expression for the stress-energy tensor of a perfect fluid. By using \(nU=J^0 u\) and the equation of state:

\begin{align*}
    T^{ab}&=p\gamma^{ab}+\frac{(J^0)^2}{n^2}(p+\rho)(\beta^a+u^a)(\beta^b+u^b)\\
    &=p\gamma^{ab}+(p+\rho)\Big(\frac{J^0}{n}\beta^a+U^a\Big)\Big(\frac{J^0}{n}\beta^b+U^b\Big)\\
\end{align*}

This is the stress-energy tensor for a perfect fluid in cosmology, shifted by the vector \(\frac{J^0}{n}\beta^a\).

\subsubsection{Constraint equations}

Since \(\alpha\) and \(\beta^a\) are dynamical variables and components of the 4-dimensional metric, they together contribute to 4 more Euler-Lagrange equations. The absence of their temporal derivatives in the action permits a Lagrange multiplier interpretation of their physical meaning. In fact, they are commonly regarded as constraint equations. The presence of the fluid action adds additional terms to the constraint equations, so the modified equations are included here.

Firstly, we compute:

\begin{align*}
    \frac{\delta\rho}{\delta \alpha}&=\frac{d\rho}{dn}\frac{\delta n}{\delta \alpha}\\
&=\frac{d\rho}{dn}\left(J^0\frac{\delta }{\delta \alpha}\sqrt{\alpha^2-\gamma_{ab}(\beta^a+u^a)(\beta^b+u^b)}+\frac{\delta J^0}{\delta \alpha}\sqrt{\alpha^2-\gamma_{ab}(\beta^a+u^a)(\beta^b+u^b)}\right)\\    
    &=\frac{d\rho}{dn} \left(\frac{(J^0)^2}{n}\alpha-\frac{n}{\alpha}\right),\\
\end{align*}
and 

\begin{align*}
    \frac{\delta\rho}{\delta \beta^a}&=\frac{d\rho}{dn}\frac{\delta n}{\delta \beta^a}\\
    &=-\frac{d\rho}{dn}\frac{(J^0)^2}{n}\gamma_{ab}(\beta^b+u^b),
\end{align*}
where we have used the fact that \(J^0 \propto \frac{1}{\alpha}\).
At last, taking variations of the action with respect to the lapse function and shift vectors yield the constraint equations \cite{alcubierre_2008_introduction}:

\begin{align*}
K_{ab}K^{ab}-K^2+R&=\rho +\alpha \frac{\delta\rho}{\delta \alpha}\\
-2D_b(K^{ab}-\gamma^{ab}K)&=\alpha \frac{\delta\rho}{\delta \beta^a},
\end{align*}
where \(D\) is the covariant derivative using the Levi-Civita connection of the metric tensor.

\subsection{Euler-Poincar\'e Reduction and Derivation of the Momentum Equation}

The principle of Euler-Poincar\'e reduction with advected quantities is summarized below. A comprehensive treatment is covered in \cite{holm_2009_geometric}.

Consider a Lagrangian \(L: TG \times V^*\to \mathbb{R}\), for some Lie group \(G\) and a vector space \(V^*\) under a representation of \(G\) acting from the right. If the Lagrangian is right-invariant under the transformation \((g,\dot{g},a_0)\to(gh,\dot{g}h,a_0h) \), the principle of least action for this Lagrangian with \(a_0\in V^*\) held as a constant parameter may be reduced by noting that:

\begin{equation} \label{redlag}
    L(g_t, \dot{g_t}, a_0)=L(e, \dot{g_t}g_t^{-1},a_0g_t^{-1} )=:l(\dot{g_t}g_t^{-1},a_0g_t^{-1} ),
\end{equation}
for some reduced Lagrangian \(l:\mathfrak{g}\times V^*\to \mathbb{R}\). The resulting variational principle is done on the reduced Lagrangian with free and holonomic variations of \(g_t\). Thus the variations of \(v:=\dot{g_t}g_t^{-1}\) and \(a:=a_0g_t^{-1} \) are:

\begin{equation}
    \delta v=\dot{\eta}-[v,\eta],\ \delta a=- a\eta,
\end{equation}
with \(\eta=\delta g_t g_t^{-1}\in \mathfrak{g}\), which is a free and holonomic variation.

The extremisation of the action yields:

\begin{align*}
0=\delta S&=\int dt\ \delta l(v,a)\\
&=\int dt\ \langle\frac{\delta l}{\delta v},\delta v\rangle+\langle\frac{\delta l}{\delta a},\delta a \rangle\\
&=\int dt\ \langle\frac{\delta l}{\delta v},\dot{\eta}-[v,\eta]\rangle+\langle\frac{\delta l}{\delta a},-a\eta \rangle\\
&=\int dt\ \langle\frac{\delta l}{\delta v},\dot{\eta}-ad_v(\eta)\rangle+\langle\frac{\delta l}{\delta a},-a\eta \rangle\\
&=\int dt\ \langle-\frac{d}{dt}\frac{\delta l}{\delta v}-ad^*_v\frac{\delta l}{\delta v},\eta\rangle+\langle\frac{\delta l}{\delta a}\diamond a,\eta\rangle.\\
\end{align*}

This implies:

\begin{equation}
    \frac{d}{dt}\frac{\delta l}{\delta v}+ad^*_v\frac{\delta l}{\delta v}=\frac{\delta l}{\delta a}\diamond a,
\end{equation}
where the diamond term \(\frac{\delta l}{\delta a}\diamond a \in \mathfrak{g}^*\) is defined such that \(\langle\frac{\delta l}{\delta a}\diamond a,\eta\rangle:=\langle\frac{\delta l}{\delta a},- a \eta\rangle\) for all \(\eta\in \mathfrak{g}\). The advection equation:

\begin{equation}
    \dot{a}=-av,
\end{equation}
also arises as the motion of \(a\) is restricted by the representation of \(G\).

This property may be applied to the present scenario with \(G=\textit{Diff}(M), \mathfrak{g}=\mathfrak{X}(M), V^*=\Lambda^3(M)\), whose elements are \((h_t, u, J^0)\) respectively. The right \(G\)-action on \(V^*\) is \( J^0h_t=h^*_tJ^0\), and the infinitesimal generator is \(J^0 u=\mathcal{L}_uJ^0\). From here on, the fluid Lagrangian is denoted as the reduced Lagrangian, \(l=-\int_M\alpha \sqrt{\gamma}\rho \ d^3x\), with only \(\rho\) varying under the Eulerian variables. It is noted that in switching to the \(3+1\) formalism, the dynamical field of the fluid has changed from \(\Psi\) to \(h_t\). Thus, for the derivation of the Euler-Poincar\'e equation and the analysis that follows, all quantities are labeled by their representations under the action of the spatial fluid diffeomorphism, \(h_t\), rather than their original, covariant descriptions. For example, \(\alpha \sqrt{\gamma} \in C^\infty(M)\) is now interpreted as a scalar since the fluid diffeomorphism acts trivially on the ADM variables. However, originally a component of a vector field, \(J^0 \in \Lambda^3(M)\) now replaces its role as the volume element due to its transformation property stated in \eqref{J^0}.

The constrained variation of the volume form \(J^0\) is given by:

\begin{equation}
    -J^0\eta =-\mathcal{L}_\eta J^0=-\partial_a(J^0 \eta^a).
\end{equation}

Then using integration by parts,

\begin{align*}
\langle\frac{\delta l}{\delta J^0}\diamond J^0,\eta\rangle&=\langle\frac{\delta l}{\delta J^0},-\mathcal{L}_\eta J^0\rangle\\
\langle\frac{\delta l}{\delta J^0}\diamond J^0,\eta\rangle&=\langle\frac{\delta l}{\delta J^0},-\partial_a(J^0 \eta^a)\rangle\\
\langle\Big(\frac{\delta l}{\delta J^0}\diamond J^0\Big)_a,\eta^a\rangle&=\langle J^0\partial_a\frac{\delta l}{\delta J^0},\eta^a\rangle.\\
\end{align*}

Therefore 

\begin{equation}
    \Big(\frac{\delta l}{\delta J^0}\diamond J^0\Big)_a=J^0\partial_a\frac{\delta l}{\delta J^0}.
\end{equation}

For \(\mathfrak{g}=\mathfrak{X}(M)\), since the Lagrangian density is a volume form, the elements of the dual Lie algebra \(\mathfrak{g}^*=\Lambda^1(M) \otimes \Lambda^3(M)\) are one-form densities, and hence \(ad^*_u=\mathcal{L}_u\). Thus, the Euler-Poincar\'e equation for self-gravitating fluids in terms of ADM variables is:

\begin{equation}
   \frac{d}{dt}\frac{\delta l}{\delta u^a}+\Big(\mathcal{L}_u\frac{\delta l}{\delta u}\Big)_a=J^0\partial_a\frac{\delta l}{\delta J^0},
\end{equation}
or equivalently, in a coordinate-free form in terms of operators,

\begin{equation}
   \Big(\frac{d}{dt}+\mathcal{L}_u\Big)\frac{\delta l}{\delta  u}=J^0d\Big(\frac{\delta l}{\delta J^0}\Big),
\end{equation}
where \(d\) is the exterior derivative on \(M\). The advection equation is:

\begin{equation}
    \dot{J^0}=-\mathcal{L}_uJ^0.
\end{equation}

Finally, explicit computation of the variational derivatives with respect to \(u\) and \(J^0\) yields:

\begin{equation}
    \frac{\delta l}{\delta u^a}=\alpha \sqrt{\gamma}\frac{d \rho}{dn}\frac{(J^0)^2}{n}\gamma_{ab}(\beta^b+u^b),
\end{equation}
and 

\begin{equation}
    \frac{\delta l}{\delta J^0}=-\alpha \sqrt{\gamma}\frac{d \rho}{dn}\frac{n}{J^0}=-\alpha \sqrt{\gamma}\frac{p+\rho}{J^0}.
\end{equation}

We note that the differential operator \(\frac{\delta}{\delta J^0}\) is performed at fixed \((\alpha, \gamma_{ab})\), as the constrained variation of \(J^0\) in the Euler-Poincar\'e equation arises from varying the fluid map \(h_t\), and not the ADM variables.

\section{Euler-Poincar\'e Equation in a Moving Frame} \label{Section 5}

In this section, we derive a variant of the action principle and the equations of motion in terms of the underlying geometry \((\gamma_{ab}, \alpha, \beta^a)\) in a fixed frame of reference, with the Eulerian variables \((u, J^0)\) measured in a separate frame of reference. Similar to expressing Newton's laws of motion in rotating frame variables, the fluid equations may be transformed by using a time-dependent coordinate transformation. We denote this as the moving frame, as the results derived here apply to any time-dependent transformation of the spatial variables.

\subsection{Transforming to Moving Coordinates}

This transformation of coordinates may be done on the Euler-Poincar\'e equation. However, the proposed variational structure may be utilised by performing the transformation on the action itself. This way, we may identify which extra terms are generated from the transformation of each variational derivative. This is done by first setting

\begin{equation}
    \Psi=O\circ \tilde{\Psi},\ O(x,t)=(O_t(x),t),
\end{equation}
for some choice of curve \(O_t\in \textit{Diff}(M)\ , \ \forall t\in \mathbb{R}\) matching the movement of an observer. The transformation acts on the fluid map only because an observer perceives Eulerian fluid variables, generated by \(\tilde{\Psi}\).

By using the invariance of the action under the gauge transformation 

\begin{equation} \label{rotgauge}
    \Psi\to O^{-1} \circ \Psi=\tilde{\Psi}, g\to O^*g,
\end{equation}
the equations of motion in a different frame of reference may be obtained by setting \(g\to O^*g\) in the total Lagrangian in \eqref{taubpullback}. We note that this gauge transformation may be done towards \((h_t,\dot{h}_t)\) instead, since \(O\) is a gauge-fixed map. The ADM Lagrangian is itself invariant under this transformation, as it does not depend on the fluid map. However, the fluid Lagrangian changes form, as shown below. 

The reference metric is decomposed as \eqref{ADMsplit}

\begin{equation}
    g=(\beta^a \beta_a -\alpha^2)dt^2+\beta_a (dx^adt+dtdx^a)+\gamma_{ab}dx^adx^b.
\end{equation}

By using the identities for the Pull-back of one-forms from Appendix \ref{Appendix A}, a direct calculation yields

\begin{align*}
O^*g=\ &\left(-\alpha^2+\gamma_{ab}(\beta^a+o^a)(\beta^b+o^b)\right)dt^2\\
&+\gamma_{bc}(\beta^b+o^b)(O_*\partial_a)^c(dtdx^a+dx^adt)\\
&+\gamma_{cd}(O_*\partial_a)^c(O_*\partial_b)^ddx^adx^b,
\end{align*}
in terms of the reference ADM variables. \(o^a\) are the components of the Eulerian velocity associated with the change of frame, \(o^a=(\dot{O}_tO_t^{-1})^a\).

It is important to note that the transformed metric evaluates \((\alpha,\beta^a,\gamma_{ab})\) at \(O(x,t)\), or simply \(O_t(x)\), by the definition of pull-back. This means that in the local expression, \(o\circ O_t=\dot{O_t}\) is also implied. \((O_*\partial_a)^c\) are elements of the block diagonal Jacobian matrix. Consequently, they may be identified as the elements of the Jacobian for the spatial transformation \(O_t\) at a certain point in time. Thus, the last term in our expression for \(O^*g\) may be expressed in terms of the pull-back metric:

\begin{equation}
    \hat{\gamma}_{ab}(x)dx^adx^b=\gamma_{cd}(O_t(x))\ (O_*\partial_a)^c(O_*\partial_b)^ddx^adx^b,
\end{equation}
or, in coordinate-free notation, \(\hat{\gamma}=O_t^*\gamma\).

Let \((\tilde{u},\tilde{J^0})\) be the Eulerian quantities perceived by the observer, associated to \(\tilde{\Psi}\) via \(\tilde{u}=\dot{\tilde{h}}_t\tilde{h}_t^{-1}\) and \(\tilde{J^0}=\frac{\tilde{n}_0}{\det(\nabla \tilde{h}_t)\alpha \sqrt{\gamma}}\). The number density is:

\begin{align*}
n(x)=&\tilde{J^0}(x)\sqrt{-O^*g(\tilde{u}(x),\tilde{u}(x))}\\
=&\tilde{J^0}(x)\Bigg[\alpha^2\Big|_{O_t(x)}-\gamma_{ab}(\beta^a+o^a)(\beta^b+o^b)\Big|_{O_t(x)}\\
&-2\gamma_{bc}(\beta^b+o^b)\Big|_{O_t(x)}\times(O_*\partial_a)^c\tilde{u}^a(x)-\hat{\gamma}_{ab}(x)\tilde{u}^a(x)\tilde{u}^b(x)\Bigg]^\frac{1}{2}.
\end{align*}

\((O_*\partial_a)^c\tilde{u}^a\) can likewise be interpreted as an artificially transformed Eulerian velocity \(\hat{u}^c:=(O_{t*}u)^c\), or in coordinates, 

\begin{equation} \label{uhat}
    \hat{u}^c(O_t(x))=(O_*\partial_a)^c\tilde{u}^a(x).
\end{equation}

Thus, all terms in the square root may be evaluated at \(O_t(x)\) using the \(\hat{u}\) notation:

\begin{equation}
    n(x)=\tilde{J^0}(x)\sqrt{\alpha^2-\gamma_{ab}(\beta^a+o^a)(\beta^b+o^b)-2\gamma_{bc}(\beta^b+o^b)\hat{u}^a-\gamma_{ab}\hat{u}^a\hat{u}^b}\Bigg|_{O_t(x)}.
\end{equation}

This is conveniently expressed as:

\begin{equation}
    n(x)=\tilde{J^0}(x)\sqrt{\alpha^2-\gamma_{ab}(\beta^a+o^a+\hat{u})(\beta^b+o^b+\hat{u})}\Bigg|_{O_t(x)}.
\end{equation}

Finally, the fluid Lagrangian takes the form:

\begin{equation}
    l=-\int_M d^3x\ \alpha \sqrt{\gamma} \rho \left( \tilde{J^0}(x)\sqrt{\alpha^2-\gamma_{ab}(\beta^a+o^a+\hat{u})(\beta^b+o^b+\hat{u})} \Bigg|_{O_t(x)}\right).
\end{equation}

It is observed that only \(\tilde{J^0}\) is evaluated at \(x\) rather than \(O_t(x)\). Now, a change of variables will be performed to remove this non-locality. We note that

\begin{equation}
    \int_M d^3x=\int_M \frac{1}{\det(\nabla O_t)} d^3(O_t(x)) ,
\end{equation}
and 

\begin{align*}
O_t^{-1*}\tilde{J^0}(x)&=\frac{1}{\det(\nabla O_t)}\tilde{J^0}(O_t^{-1}(x))\\
\tilde{J^0}(x)&=\det(\nabla O_t)\ O_t^{-1*}\tilde{J^0}(O_t(x)).
\end{align*}

Define 

\begin{equation}
    \tilde{\rho}(n)=\frac{1}{\det(\nabla O_t)}\rho(\det(\nabla O_t)n),
\end{equation}
and

\begin{equation} \label{J0hat}
    \hat{J^0}=O_t^{-1*}\tilde{J^0}.
\end{equation}

The Lagrangian after a change of variables is:

\begin{equation}
    l=-\int_M d^3x\ \alpha \sqrt{\gamma} \tilde{\rho} \left( \hat{J^0}\sqrt{\alpha^2-\gamma_{ab}(\beta^a+o^a+\hat{u})(\beta^b+o^b+\hat{u})} \right).
\end{equation}

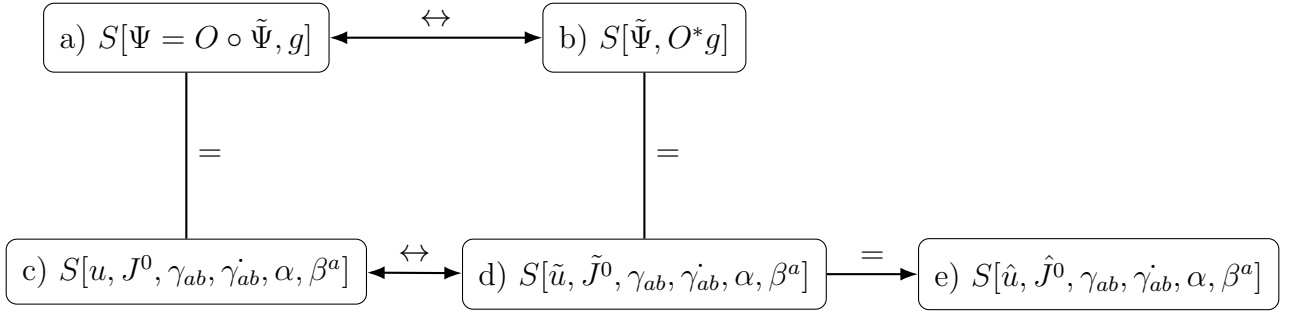
\begin{figure}[t]
\centering
\begin{tikzpicture}[
  box/.style={draw, rounded corners, inner sep=6pt, align=center, minimum width=10mm},
  dblarr/.style={<->, >=Latex, thick},
  eqarr/.style={-, thick}
]

\node[box] (a) {a) $S[\Psi=O\circ\tilde{\Psi},g]$};
\node[box, right=28mm of a] (b) {b) $S[\tilde{\Psi},O^*g]$};

\node[box, below=22mm of a] (c) {c) $S[u,J^0, \gamma_{ab}, \dot{\gamma_{ab}},\alpha,\beta^a]$};
\node[box, below=22mm of b] (d) {d) $S[\tilde{u},\tilde{J^0}, \gamma_{ab}, \dot{\gamma_{ab}},\alpha,\beta^a]$};

\node[box, right=12mm of d] (e) {e) $S[\hat{u},\hat{J^0}, \gamma_{ab}, \dot{\gamma_{ab}},\alpha,\beta^a]$};

\node[box, align=center, font=\bfseries, above=8mm of a]
  {Covariant formalisms: a), b)};
\node[box, align=center, font=\bfseries, below=10mm of c]
  {$3+1$ formalisms: c), d), e)};

\draw[dblarr] (a) -- node[above] {$\leftrightarrow$} (b);

\draw[eqarr] (a) -- node[right] {$=$} (c);
\draw[eqarr] (b) -- node[right] {$=$} (d);

\draw[dblarr] (c) -- node[above] {$\leftrightarrow$} (d);
\draw[->, >=Latex, thick] (d) -- node[above] {$=$} (e);

\end{tikzpicture}
\caption{\footnotesize Schematic representation of equivalent variational principles of a general relativistic barotopic fluid. a) and b) are covariant, field-theoretic descriptions with actions in the form of \eqref{taubpullback} using the fluid map and metric, whereas c), d), and e) are \(3+1\) formalisms with actions in the form of \eqref{S3+1}, using ADM variables for the metric, and the Eulerian structure highlighted in \eqref{redlag} in Section \ref{Section 4} to describe hydrodynamics. "\(=\)" signs connect a) to c) \eqref{S}, b) to d), and d) to e) \eqref{uhat} \eqref{J0hat}, meaning they are related by a change of variables. "\(\leftrightarrow\)" signs connect a) to c) and b) to d), as they are related by spacetime gauge transformations, as shown in \eqref{rotgauge}.}
\label{fig2}
\end{figure}

In this section, we have introduced three new representations of the system to facilitate descriptions in the moving frame of reference. Figure \ref{fig2} reviews the steps taken in this calculation.

\subsection{Stress-Energy Tensor and Euler-Poincar\'e Equation}

Since \(O\) is a choice of reference frame rather than a dynamical variable, and since the variations of \((h_t,\dot{h}_t)\to(\tilde{h}_t, \dot{\tilde{h}}_t)\) remain free and holonomic, the geometric structure in terms of the new variables is unchanged, and the Euler-Poincar\'e reduction may be applied to the perceived Eulerian variables \((\tilde{u},\tilde{J^0})\). The modifications to the equations of motion arise from the changes to the variational derivatives, which are all computed below.

\begin{align*}
\frac{\delta l}{\delta \alpha}&=-\sqrt{\gamma}\tilde{\rho} -\alpha \sqrt{\gamma}\frac{\delta\tilde{\rho}}{\delta \alpha}\\
&=-\sqrt{\gamma}\tilde{\rho}-\alpha \sqrt{\gamma}\frac{d\tilde{\rho}}{dn} \Big(\frac{(\hat{J^0})^2}{n}\alpha -\frac{n}{\alpha}\Big)\\
&=\sqrt{\gamma}\tilde{p}-\alpha^2\sqrt{\gamma}\frac{d\tilde{\rho}}{dn} \frac{(\hat{J^0})^2}{n},
\end{align*}
with a new pressure quantity

\begin{equation}
    \tilde{p}=n\frac{d\tilde{\rho}}{dn}-\tilde{\rho}.
\end{equation}

Other than the renaming of variables, the variational derivative for the lapse function takes the same functional form as before. In addition to that, the variational derivatives with respect to the shift vector and the spatial metric are shifted by \(o^a\) as follows:

\begin{align*}
\frac{\delta l}{\delta \beta^a}=-\alpha \sqrt{\gamma}\frac{d\tilde{\rho}}{dn}\frac{\delta n}{\delta \beta^a}&=\alpha \sqrt{\gamma}\frac{d\tilde{\rho}}{dn}\frac{(\hat{J^0})^2}{n}\gamma_{ab}(\beta^b+\hat{u}^b+o^b),
\end{align*}
and
\begin{equation}
    \frac{\delta l}{\delta \gamma_{ab}}=\frac{1}{2}\alpha \sqrt{\gamma}\tilde{p} \gamma^{ab}+\frac{1}{2}\frac{(\hat{J^0})^2}{n}\alpha \sqrt{\gamma}\frac{d\tilde{\rho}}{dn}(\beta^a+\hat{u}^a+o^a)(\beta^b+\hat{u}^b+o^b).
\end{equation}

Thus, the stress-energy tensor is:

\begin{equation}
    T^{ab}=\tilde{p}\gamma^{ab}+\frac{(\hat{J^0})^2}{n}\frac{d\tilde{\rho}}{dn}(\beta^a+\hat{u}^a+o^a)(\beta^b+\hat{u}^b+o^b).
\end{equation}

While Einstein's equations remain mostly the same under this change of coordinates, this is not the case for the Euler-Poincar\'e equation. Although the perceived Eulerian variables satisfy the Euler-Poincar\'e equation

\begin{equation} \label{EPfortilde}
   \Big(\frac{d}{dt}+\mathcal{L}_{\tilde{u}}\Big)\frac{\delta l}{\delta  \tilde{u}}=\tilde{J^0}d\Big(\frac{\delta l}{\delta \tilde{J^0}}\Big),
\end{equation}
as well as the advection equation

\begin{equation}
    \dot{\tilde{J^0}}=-\mathcal{L}_{\tilde{u}}\tilde{J^0},
\end{equation}
finding the variational derivatives \(\frac{\delta l}{\delta  \tilde{u}}\) and \(\frac{\delta l}{\delta \tilde{J^0}}\) is nontrivial, as they do not appear explicitly in the Lagrangian. To illustrate this, we note that \(\hat{u}\) is not the true Eulerian velocity of the fluid, nor is \(\hat{J^0}\) the true advected quantity. They are only shorthand for the push-forwards of the perceived velocity and density. The variables to differentiate with respect to are thus \(\tilde{u}\) and \(\tilde{J^0}\) rather than \(\hat{u}\) and \(\hat{J^0}\), and the chain rule must be used to obtain the Euler-Poincar\'e equation explicitly. The variational derivative with respect to \(\hat{u}\) is:

\begin{equation}
    \frac{\delta l}{\delta \hat{u}^a}=\alpha \sqrt{\gamma}\frac{d \tilde{\rho}}{dn}\frac{(\hat{J^0})^2}{n}\gamma_{ab}(\beta^b+\hat{u}^b+o^b).
\end{equation}

Recalling \eqref{uhat}, and by making use of the chain rule, we have:

\begin{align*}
\frac{\delta l}{\delta\tilde{u}^a(x)}&=\int_M dx'^3\frac{\delta l}{\delta \hat{u}^c(x')} \frac{\delta \hat{u}^c(x')}{\delta \tilde{u}^a(x)}\\
&=\int_M dx'^3\frac{\delta l}{\delta \hat{u}^c(x')}(O_*\partial_a)^c \delta^3(x'-O_t(x))\\
&=\frac{\delta l}{\delta \hat{u}^c(O_t(x))}(O_*\partial_a)^c\\
&=\frac{\delta l}{\delta \hat{u}^c}(O_*\partial_a)^c\Bigg|_{O_t(x)}.\\
\end{align*}

It is recalled that \(\frac{\delta l}{\delta\hat{u}}\in \Lambda^1(M) \otimes\Lambda^3(M)\). The pull-back of this quantity is

\begin{equation}
     O_t^*(\frac{\delta l}{\delta \hat{u}})\Bigg|_x=\frac{\delta l}{\delta \hat{u}^c}(O_*\partial_a)^c\det(\nabla O_t)\Bigg|_{O_t(x)}.
\end{equation}

Thus, the variational derivative with respect to the perceived Eulerian velocity may be understood in a coordinate-free form:

\begin{align*}
    \frac{\delta l}{\delta \tilde{u}}&=\frac{1}{\det(\nabla O_t)} O_t^*\Big(\frac{\delta l}{\delta \hat{u}}\Big)\\
    &=\frac{1}{\det(\nabla O_t)} O_t^*\left(\alpha \sqrt{\gamma}\frac{d \tilde{\rho}}{dn}\frac{(\hat{J^0})^2}{n}\gamma_{ab}(\beta^b+\hat{u}^b+o^b)dx^a\right).
\end{align*}

As for \(\tilde{J^0}\),

\begin{equation}
    \frac{\delta l}{\delta \hat{J^0}}=-\alpha \sqrt{\gamma}\frac{\tilde{p}+ \tilde{\rho}}{\hat{J^0}}.
\end{equation}

Recalling \eqref{J0hat}, and by using the chain rule once again,
\begin{equation}
    \hat{J^0}(O_t(x))=\frac{1}{\det(\nabla O_t)}\tilde{J^0}(x),
\end{equation}

\begin{align*}
    \frac{\delta l}{\delta \tilde{J^0}(x)}&=\int_M dx'^3\frac{\delta l}{\delta \hat{J^0}(x')} \frac{\delta \hat{J^0}(x')}{\delta \tilde{J^0}(x)}\\
    &=\int_M dx'^3\frac{\delta l}{\delta \hat{J^0}(x')}\frac{1}{\det(\nabla O_t)}\delta^3(x'-O_t(x)) \\
    &=\frac{\delta l}{\delta \hat{J^0}}\frac{1}{\det(\nabla O_t)}\Bigg|_{O_t(x)}.
\end{align*}

Furthermore, by noting that \(\frac{\delta l}{\delta \hat{J^0}}\), being dual to \(\hat{J^0}\), transforms as a scalar:

\begin{equation}
    O_t^*\left(\frac{\delta l}{\delta \hat{J^0}}\right)\Bigg|_x=\frac{\delta l}{\delta \hat{J^0}}\Bigg|_{O_t(x)}.
\end{equation}

Therefore, the variational derivative for \(\tilde{J^0}\) is:
\begin{align*}
    \frac{\delta l}{\delta \tilde{J^0}}&=\frac{1}{\det(\nabla O_t)}O_t^*\left(\frac{\delta l}{\delta \hat{J^0}}\right)\\
    &=\frac{1}{\det(\nabla O_t)}O_t^*\left(-\alpha \sqrt{\gamma}\frac{\tilde{p}+ \tilde{\rho}}{\hat{J^0}}\right).
\end{align*}

Finally, the Euler-Poincar\'e equation in any frame of reference can be written in the coordinate-free form:

\begin{equation}
   \Big(\frac{d}{dt}+\mathcal{L}_{\tilde{u}}\Big)\left(\frac{1}{\det(\nabla O_t)}O_t^*(\frac{\delta l}{\delta \hat{u}})\right)=\tilde{J^0}d\left(\frac{1}{\det(\nabla O_t)}O_t^*\left(\frac{\delta l}{\delta \hat{J^0}}\right)\right),
\end{equation}
and the advection equation is:

\begin{equation}
    \dot{\tilde{J^0}}=-\mathcal{L}_{\tilde{u}}\tilde{J^0}.
\end{equation}

It is important to note that the Lie derivatives are still done with respect to \(\tilde{u}\), the perceived Eulerian velocity, and that \(\tilde{J^0}\) now lives in a representation of \(\tilde{h}_t\) instead of \(h_t\). This calculation may be done abstractly as in \eqref{EPfortilde}, but it is placed in a form that is viable for computational simulation.

\section{Kelvin Circulation Theorem} \label{Section 6}

In Section \ref{Section 3}, we have mentioned that no Noether charges arise from continuous symmetry in the Pull-back Action, as we have assumed spacetime to have no boundaries. However, this section shows that conserved quantities arise in the fluid system, as expected from the Euler-Poincar\'e theory \cite{holm_1998_the}. This is the circulation integral, a quantity which may be evaluated along any spatial loop that follows the motion of the fluid.

First, it is noted that the Euler-Poincar\'e equation concerns one-form densities \(\Lambda^1(M) \otimes\Lambda^3(M)\). Dividing by the advected volume form \(J^0\) yields an equation for one-forms:

\begin{equation}
   \frac{1}{J^0}\Big(\frac{d}{dt}+\mathcal{L}_u\Big)\frac{\delta l}{\delta  u}=d\Big(\frac{\delta l}{\delta J^0}\Big).
\end{equation}

Then, by using the advection equation, it is seen that \(J^0\) is a constant under the differential operator \((\frac{d}{dt}+\mathcal{L}_u)\) when following the fluid flow. Thus,

\begin{equation}
  \Big(\frac{d}{dt}+\mathcal{L}_u\Big) \left( \frac{1}{J^0}\frac{\delta l}{\delta  u}\right)=d(\frac{\delta l}{\delta J^0}).
\end{equation}

Now, by considering any closed loop \(C\) on \(M\) flowing in time according to the Eulerian velocity \(u\), integrating the above equation along that loop yields:

\begin{equation} \label{kelvincirc}
    \frac{d}{dt}\oint_C  \frac{1}{J^0}\frac{\delta l}{\delta  u}=\oint_C \Big(\frac{d}{dt}+\mathcal{L}_u\Big) \left( \frac{1}{J^0}\frac{\delta l}{\delta  u}\right)=\oint_C d \Big(\frac{\delta l}{\delta J^0}\Big)=0.
\end{equation}

In going from the first to the second expression, the Reynolds transport theorem for loop integrals is used. The second and third expressions are related by the Euler-Poincar\'e equation. The final expression vanishes, as the integral of an exact differential 1-form along a closed loop yields zero by the fundamental theorem of calculus.

The integral may be interpreted as the circulation around the given loop \(C\), which is being advected by the fluid flow. Its conservation is the Kelvin circulation theorem for the Euler-Poincar\'e equation. By writing the variational derivative in terms of the dynamical quantities, we obtain the following:

\begin{equation}
    \frac{d}{dt}\oint_C \alpha \sqrt{\gamma}\frac{d \rho}{dn}\frac{J^0}{n}\gamma_{ab}(\beta^b+u^b) dx^a=0.
\end{equation}

A similar argument may be written for the Euler-Poincar\'e equation in a moving frame. By considering any loop \(\tilde{C}\) advected by \(\tilde{u}\) instead, dividing the Euler-Poincar\'e equation in the moving frame by \(\tilde{J^0}\) yields:

\begin{equation}
    \frac{1}{\tilde{J^0}}(\frac{d}{dt}+\mathcal{L}_{\tilde{u}})\left(\frac{1}{\det(\nabla O_t)}O_t^*(\frac{\delta l}{\delta \hat{u}})\right)=d\left(\frac{1}{\det(\nabla O_t)}O_t^*\left(\frac{\delta l}{\delta \hat{J^0}}\right)\right).
\end{equation}

Following the same procedure as the inertial case in \eqref{kelvincirc}, the conserved loop integral is:

\begin{equation}
    \frac{d}{dt}\oint_{\tilde{C}} \frac{1}{\tilde{J^0}}\frac{1}{\det(\nabla O_t)}O_t^*(\frac{\delta l}{\delta \hat{u}})=0.
\end{equation}

Using \eqref{J0hat}, we obtain

\begin{equation}
    \frac{d}{dt}\oint_{\tilde{C}} \frac{1}{\det(\nabla O_t)}O_t^*\left(\frac{1}{\hat{J^0}}\frac{\delta l}{\delta \hat{u}}\right)=0.
\end{equation}

Thus, the Kelvin circulation theorem is:

\begin{equation}
    \frac{d}{dt}\oint_{\tilde{C}} \frac{1}{\det(\nabla O_t)}O_t^*\left(\alpha \sqrt{\gamma}\frac{d \tilde{\rho}}{dn}\frac{\hat{J^0}}{n}\gamma_{ab}(\beta^b+\hat{u}^b+o^b)dx^a\right)=0.
\end{equation}

It can be seen that the loop integral above reverts back to the circulation integral for fluids in an inertial frame, by setting \(O=id_W\). 

In this section, we have shown that the motion of the fluid is circulation-preserving. In addition, this structure is exhibited in all frames of reference for the fluid variables relative to gravity. We recall that the spacetime gauge transformation of the Pull-back action acts on the gravity and fluid variables simultaneously, and that the fluid does not possess gauge invariance relative to gravity. However, we have shown that circulation-preservation is a retained structure in this frame. This is because the system has an Euler-Poincar\'e structure across all frames of reference due to gauge invariance, as shown in Section \ref{Section 5}.

\section{Conclusions}

To summarise, Section \ref{Section 3} showed that by performing the \(3+1\) decomposition of spacetime on the Pull-back Action for relativistic hydrodynamics, it not only permits the trivialisation of the metric in terms of ADM variables, but also trivialises the fluid map by using gauge invariance. This permits an explicit expression of the action in terms of the spatial fluid diffeomorphism, turning the covariant action into regular Lagrangian mechanics, and setting the stage for Euler-Poincar\'e reduction. The Euler-Poincar\'e equations for barotropic fluids were derived in fixed and moving frames of reference, followed by derivations of the corresponding circulation theorems.

Immediate applications of these equations are limited due to the standardised usage of high-resolution shock capturing methods, requiring that the equations be put into hyperbolic form, such as the Valencia formulation \cite{mart_1991_numerical}. However, Section \ref{Section 4} opens up new prospects for cross-disciplinary approaches by placing the equations of motion in a form that parallels nonrelativistic fluid dynamics, potentially allowing methods from computational fluid dynamics to be transferred here. In addition, Section \ref{Section 5} derives the equations of motion in a moving frame of reference, which may be used to remove the predicted bulk motion of the fluid in simulations. This reduces advection-related errors and drifts in density terms \cite{duez_2003_hydrodynamic}. Finally, circulation for general relativistic fluids is defined and proven to be conserved in this paper, offering room for analysis and the development of techniques regarding its preservation.

Future extensions to this may include general relativistic magnetohydrodynamics, fluids with thermal properties and advection, or multifluid systems. The above variational formalism offers a foundation for the formal derivation of the equations by encoding the kinematic structure and group actions of the dynamical fields, potentially by modelling a covariant Lagrangian for a multifluid system using a semidirect product of diffeomorphisms \(\textit{Diff}(W)\ltimes \textit{Diff}(W)\). The circulation theorem of the system may be derived using the methods highlighted in Section \ref{Section 6}, from the corresponding Euler-Poincar\'e equations.

\section*{Acknowledgements}

I am grateful to Darryl Holm for the inspiration for this project, and for the meaningful discussions during the course of this work, which have refined the interpretation of its results. I would also like to thank Logan Arnold-Lee, Calvin Lim, Jordan Pefianco, and Inesh Muhkerjee for their thoughtful suggestions for the manuscript.

\appendix 

\section{Transformation Rules Under Gauge-Fixed Maps} \label{Appendix A}

In this Appendix, we derive the transformation rules of geometrical objects on a trivial bundle \(W=M \times \mathbb{R}\) under the map:

\begin{equation}
    \Psi(x,t)=(h_t(x),t).
\end{equation}

The following identities are used in Appendix \ref{Appendix B} when calculating the number density current explicitly, and in Section \ref{Section 5} when finding a coordinate expression for the pull-back metric.

\subsection{Basis vector fields}

The tangent basis of \(W\) is \(\{\partial_1,\partial_2, \partial_3, \partial_t\} \), with \(\partial_t\) the vertical direction. The rules for the push-forwards under \(\Psi\) are:
\begin{equation}
    \Psi_* \partial_t= \partial_t+ \dot{h}_th_t^{-1},
\end{equation}
\begin{equation}
    \Psi^{-1}_* \partial_t= \partial_t- \dot{h}_th_t^{-1},
\end{equation}
\begin{equation}
    \Psi_*\partial_a=h_{t*} \partial_a,
\end{equation}
with \(a=\{1,2,3\}\), and \(h_{t*}\) is the analog of the push-forward map on tangent vectors on the base space \(M\). We also note that the above implies

\begin{equation}
     \Psi_* (\dot{h}_th^{-1}_t)=(\dot{h}_th^{-1}_t),
\end{equation}
with \(\dot{h}_th^{-1}_t \in \mathfrak{X}(M)\) a purely horizontal vector field extended to \(W\), and is in the Lie algebra of \(\textit{Diff}(M)\), since

\begin{align*}
  \partial_t&=\Psi^{-1}_*(\Psi_* \partial_t)\\
  &= \Psi^{-1}_*(\partial_t+ \dot{h}_th_t^{-1})\\
  &=\Psi^{-1}_*\partial_t+\Psi^{-1}_*(\dot{h}_th_t^{-1})\\
  &=\partial_t- \dot{h}_th_t^{-1}+\Psi^{-1}_*(\dot{h}_th_t^{-1}),
\end{align*}
implying that

\begin{align*}
 \dot{h}_th_t^{-1}&=\Psi^{-1}_*(\dot{h}_th_t^{-1})\\
 \Psi_*(\dot{h}_th_t^{-1})&=\dot{h}_th_t^{-1}.
\end{align*}

\subsection{One-Forms}

The cotangent basis of \(W\) is \(\{dx^1, dx^2, dx^3, dt\}\). The rules for the pull-backs under \(\Psi\) are as follows:

\begin{align*}
\Psi^* dt (\partial_t)&= dt(\Psi_* \partial_t)\\
&=dt(\partial_t+ \dot{h}_th_t^{-1})=1,
\end{align*}

\begin{align*}
\Psi^* dt (\partial_a)&= dt(\Psi_* \partial_a)\\
&=dt(h_{t*} \partial_a)=0.
\end{align*}

Thus \(\Psi^* dt=dt\). As for the horizontal directions:

\begin{align*}
\Psi^* dx^a (\partial_t)&=dx^a(\Psi_*\partial_t)\\
&=dx^a(\partial_t+ \dot{h}_th_t^{-1})\\
&=( \dot{h}_th_t^{-1})^a,
\end{align*}

\begin{align*}
\Psi^* dx^a (\partial_b)&=dx^a(\Psi_*\partial_b)\\
&=(\Psi_*\partial_b)^a.
\end{align*}

Thus,

\begin{equation}
    \Psi^* dx^a=(\dot{h}_th_t^{-1})^a dt+(\Psi_*\partial_b)^a dx^b.
\end{equation}

It is noted that for all calculations,

\begin{equation}
    (\Psi_*\partial_b)^a \equiv \frac{\partial x^{'a}}{\partial x^b}, \ x'=\Psi(x),
\end{equation}
is the notation used in this paper for the Jacobian matrix, in contrast with standard literature in general relativity.

\subsection{Volume forms}

Consider \(\rho=\rho_0 d^4x \in \Lambda^4(W)\).

\begin{align*}
\Psi^* \rho&=\Psi^*(\rho_0 dt\wedge dx^1 \wedge dx^2 \wedge dx^3)\\
&=\rho_0 dt\wedge  \Psi^*(dx^1 \wedge dx^2 \wedge dx^3)\\
&=\rho_0 dt\wedge  \left((\Psi_*\partial_a)^1dx^a \wedge (\Psi_*\partial_b)^2dx^b \wedge (\Psi_*\partial_c)^3dx^c\right) \\
&=\rho_0 dt\wedge  \left( \det(\nabla h_t) dx^1 \wedge dx^2 \wedge dx^3\right)\\
&=\det(\nabla h_t) \rho_0 dt\wedge dx^1 \wedge dx^2 \wedge dx^3\\
&=\det(\nabla h_t) \rho,
\end{align*}
where, in going from the second to the third line, we used the fact that \(dt\wedge dt=0\), so that only the horizontal components of the basis co-vectors contribute. \(\det(\nabla h_t)\) is the determinant of the Jacobian matrix for the horizontal map only, meaning that \(\Psi^* \rho\) does not depend on \(\dot{h}_t\).

\section{Explicit Calculation of the Number Density Current} \label{Appendix B}

In this Appendix, we express the number density current in terms of the gauge-fixed fluid map. The result is used in Section \ref{Section 3} when decomposing the Pull-back Lagrangian to \(3+1\) dimensions.

A local expression for equation\eqref{J2.2} is:

\begin{equation} \label{J}
    \tilde{n}(\pi_* V,\pi_* W,\pi_* U)=\epsilon_g (J,\Psi_* V, \Psi_* W, \Psi_* U),
\end{equation}
for all tangent vectors \(V,W,U \in T_wW, \forall w\in W\).

Let \(V=\partial_1, W=\partial_2,U=\partial_3\). Since \(\pi_* \partial_a=\partial_a\), and since \(\Psi_*\partial_a=h_{t*} \partial_a\) is purely horizontal, using the identities derived in Appendix \ref{Appendix A}:

\begin{align*}
\tilde{n}(\partial_1, \partial_2, \partial_3)&=\epsilon_g (J,\Psi_* \partial_1, \Psi_* \partial_2, \Psi_* \partial_3)\\
\tilde{n}_0&=\epsilon_g (J^0 \partial_t,\Psi_* \partial_1, \Psi_* \partial_2, \Psi_* \partial_3)\\
\tilde{n}_0&=\Psi^*\epsilon_g (J^0 \Psi^{-1}_*\partial_t, \partial_1, \partial_2,  \partial_3)\\
\tilde{n}_0&=J^0\det(\nabla h_t)\epsilon_g ( (\partial_t- \dot{h}_th_t^{-1}), \partial_1, \partial_2,  \partial_3)\\
\tilde{n}_0&=J^0 \det(\nabla h_t)\epsilon_g (\partial_t, \partial_1, \partial_2,  \partial_3)\\
\tilde{n}_0&=J^0 \det(\nabla h_t)\alpha \sqrt{\gamma}.\\
\end{align*}

Thus,

\begin{equation}
    J^0=\frac{\tilde{n}_0}{\det(\nabla h_t)\alpha \sqrt{\gamma}},
\end{equation}
with \(\tilde{n}_0\) defined as the coefficient of the fluid label volume form. In going from the fourth to fifth line, it is noted that the vectors in the other three components are horizontal, so that only the vertical component of \(\partial_t- \dot{h}_th_t^{-1}\) contributes.

Now take \(V=\partial_1, W=\partial_2,U=\partial_t\). Since \(\pi_*\partial_t=0\), the left-hand side of \eqref{J} equates to zero.

\begin{align*}
0&=\epsilon_g (J,\Psi_* \partial_1, \Psi_* \partial_2, \Psi_* \partial_t)\\
&=\epsilon_g (J,\Psi_* \partial_1, \Psi_* \partial_2, \partial_t+ \dot{h}_th_t^{-1})\\
&=\epsilon_g (J,\Psi_* \partial_1, \Psi_* \partial_2, \partial_t)+\epsilon_g (J,\Psi_* \partial_1, \Psi_* \partial_2, \dot{h}_th_t^{-1})\\
&=J^a\epsilon_g (\partial_a,\Psi_* \partial_1, \Psi_* \partial_2, \partial_t)+J^0\epsilon_g (\partial_t,\Psi_* \partial_1, \Psi_* \partial_2, \dot{h}_th_t^{-1})\\
&=J^a\Psi^*\epsilon_g (\Psi^{-1}_*\partial_a,\partial_1, \partial_2, \Psi^{-1}_*\partial_t)+J^0\Psi^*\epsilon_g (\Psi^{-1}_*\partial_t, \partial_1, \partial_2, \Psi^{-1}_*(\dot{h}_th_t^{-1}))\\
&=J^a\det(\nabla h_t)\epsilon_g (\Psi^{-1}_*\partial_a,\partial_1, \partial_2, \partial_t- \dot{h}_th_t^{-1})+J^0\det(\nabla h_t)\epsilon_g (\partial_t- \dot{h}_th_t^{-1},\partial_1, \partial_2, \dot{h}_th_t^{-1})\\
&=J^a\det(\nabla h_t)\epsilon_g (\Psi^{-1}_*\partial_a,\partial_1, \partial_2, \partial_t)+J^0\det(\nabla h_t)\epsilon_g (\partial_t,\partial_1, \partial_2, \dot{h}_th_t^{-1}).\\
\end{align*}

Simplifying, the last line yields:

\begin{align*}
J^a\epsilon_g (\Psi^{-1}_*\partial_a,\partial_1, \partial_2, \partial_t)&=-J^0\epsilon_g (\partial_t,\partial_1, \partial_2, \dot{h}_th_t^{-1})\\
-J^a\epsilon_g (\partial_t,\partial_1, \partial_2, \Psi^{-1}_*\partial_a)&=-J^0\epsilon_g (\partial_t,\partial_1, \partial_2, \dot{h}_th_t^{-1})\\
-J^a\alpha \sqrt{\gamma} ( \Psi^{-1}_*\partial_a)^3&=-J^0\alpha \sqrt{\gamma}( \dot{h}_th_t^{-1})^3\\
J^a ( \Psi^{-1}_*\partial_a)^3&=J^0( \dot{h}_th_t^{-1})^3.\\
\end{align*}

The equation above was obtained by choosing \(V,W,U\) to span the basis vector fields with \(\partial_3\) missing. Choices with \(\partial_1\) or \(\partial_2\) missing yield similar equations.

Thus,

\begin{align*}
J^a  \Psi^{-1}_*\partial_a&=J^0 \dot{h}_th_t^{-1}\\
J^a \partial_a&=J^0 \Psi_*(\dot{h}_th_t^{-1})\\
J^a  \partial_a&=J^0 \dot{h}_th_t^{-1},\\
\end{align*}
which is the equation for the 3-dimensional Eulerian velocity associated with \(h_t\).

\smallskip

\bibliographystyle{unsrt}
\bibliography{EPGR_final_ref}
\end{document}